\documentclass[twocolumn, tighten, times, twocolappendix]{aastex631}

\begin{document}

\title{Validation of {\it Gaia} DR3 orbital and acceleration solutions with hierarchical triples}

\author[0000-0002-1386-0603]{Pranav Nagarajan}
\affiliation{Department of Astronomy, California Institute of Technology, 1200 E. California Blvd., Pasadena, CA 91125, USA}

\author[0000-0002-6871-1752]{Kareem El-Badry}
\affiliation{Department of Astronomy, California Institute of Technology, 1200 E. California Blvd., Pasadena, CA 91125, USA}



\begin{abstract}

Using data from {\it Gaia} DR3, we construct a sample of 14,791 gravitationally bound wide pairs in which one of the components is an unresolved binary with an astrometric orbital or acceleration solution. These systems are hierarchical triples, with inner binary separations of order $1$ au, and outer separations of $100$--$100,000$ au. Leveraging the fact that the inner binary and outer tertiary should have nearly identical parallaxes, we use the sample to calibrate the parallax uncertainties for orbital and acceleration binary solutions. We find that the parallax uncertainties of orbital solutions are typically underestimated by a factor of $1.3$ at $G > 14$, and by a factor of $1.7$ at $G = 8$--$14$. The true parallax uncertainties are nevertheless a factor of $\sim 10$ smaller than those of the single-star astrometric solutions for the same sources. The parallax uncertainties of acceleration solutions are underestimated by larger factors of $2$--$3$ but still represent a significant improvement compared to the sources' single-star solutions. We provide tabulated uncertainty inflation factors for astrometric binary solutions and make the catalog of hierarchical triples publicly available. 

\end{abstract}



\section{Introduction} \label{sec:intro}

In recent years, the \textit{Gaia} mission \citep{gaia_mission} has dramatically increased the number of sources with precise parallax and proper motion measurements. Most of these sources display motion on the plane of the sky that is well-described by a single-star, or $5$-parameter, solution. However, many other sources exhibit astrometric ``wobble'' that cannot be satisfactorily explained by a single-star model, implying either photometric variability or binarity \citep{halbwachs_2023}. Indeed, binaries are common --- about half of all solar-type stars have at least one companion \citep[][]{moe_di_stefano_2017}. To fit the astrometry of binaries with detectable photocenter wobble, \textit{Gaia}'s third data release \citep[DR3;][]{gaia_dr3, arenou_2023} introduced new types of astrometric solutions for binaries, including both acceleration solutions designed for partial orbits and full Keplerian orbital solutions. 

\textit{Gaia} astrometric solutions hold exciting promise for the discovery of dormant stellar-mass black holes, exoplanets, rare binaries, and more \citep[][]{2024NewAR..9801694E}. However, the reliability of these solutions and their uncertainties is difficult to assess because \textit{Gaia} is much more sensitive than previous missions. To address this issue, we take advantage of the fact that many binaries with astrometric orbital or acceleration solutions have spatially resolved, gravitationally bound companions at wide separations. These systems are hierarchical triples, with an unresolved binary being orbited by a distant tertiary companion. Since the unresolved binary and the resolved tertiary are expected to have essentially the same parallax, hierarchical triples provide a useful opportunity to validate the parallaxes of astrometric binary solutions in the \textit{Gaia} DR3 catalog.

In \citet{million_binaries} (henceforth ERH21), the authors used \textit{Gaia} eDR3 \citep{gaia_edr3} data to construct a sample of spatially resolved binaries within $1$ kpc, searching out to wide separations ($s \lesssim 10^5$ au) and faint magnitudes ($G \lesssim 20.7$). By empirically estimating chance alignment probabilities, they selected pure subsets of the catalog containing only gravitationally bound pairs. They found that \textit{Gaia} eDR3 parallax uncertainties were generally reliable for faint stars ($G \gtrsim 18$), but underestimated by up to $30\%$ for brighter stars. They also found that uncertainties were more severely underestimated (up to $\sim 80\%$) for sources with companions within $4$ arcseconds, whose astrometry is disturbed by blended light from the companion. They developed a fitting function to inflate parallax uncertainties for isolated sources as a function of $G$-band apparent magnitude.

Having only been introduced in \textit{Gaia} DR3, astrometric binary solutions were not available to ERH21 and should not be expected to have the same parallax underestimate factors as single star solutions. Radial velocity follow-up of individual binaries with astrometric solutions in \textit{Gaia} DR3 has suggested that the uncertainties on the parameters of these binaries are also underestimated in at least some cases \citep[e.g.][]{chakrabarti_2023, gaia_bh1_2024, gaia_ns1}. In this study, we extend the analysis of ERH21 to wide triples containing unresolved astrometric binaries with orbital or acceleration solutions in \textit{Gaia} DR3. About half of these sources were rejected from the sample curated by ERH21 because the parallaxes from their $5$-parameter solutions were not consistent with the parallaxes of their resolved tertiary companions. The remainder were included in the ERH21 sample, but were not recognized as being part of hierarchical triples because only their $5$-parameter solutions were available at the time.

The remainder of the paper is organized as follows. In Section \ref{sec:sample}, we describe how our sample of triples hosting unresolved binaries with orbital or acceleration solutions is selected and cleaned. In Section \ref{sec:calibration}, we use the sample to validate the parallax uncertainties for these systems in \textit{Gaia} DR3. Finally, we discuss our results and summarize our main findings in Section \ref{sec:summary}.

\section{Sample Selection} \label{sec:sample}

The initial sample of candidates is selected using a procedure very similar to the one described in ERH21, but modified to use astrometric binary solutions rather than single-star solutions when these are available. In summary, all sources are retrieved from the \texttt{gaiadr3.gaia\_source} table in the \textit{Gaia} archive that have parallaxes greater than 1 mas, fractional parallax uncertainties less than 20\%, absolute parallax uncertainties less than 2 mas, and published $G$-band magnitudes. Then, the parallaxes, positions, and proper motions for sources that have astrometric solutions in the \texttt{gaiadr3.nss\_two\_body\_orbit} or \texttt{gaiadr3.nss\_acceleration\_astro} catalogs are replaced with their values from the respective catalogs. We apply the zeropoint correction of \citet{Lindegren_2021_zp} to all parallaxes, and inflate the parallax uncertainties of the tertiaries according to the fitting function $f(G)$ given in Equation 16 of ERH21. Next, pairs of sources with (1) projected separation less than 1 parsec \citep[i.e.\ the separation beyond which the Galactic tidal field rapidly disrupts binaries, see e.g.][] {2008gady.book.....B}, (2) component parallaxes consistent within 4 (or 8) sigma for angular separation greater than (less than) 4 arcseconds, and (3) component proper motions consistent within $2\sigma$ with a Keplerian orbit \citep[following][]{elbadry_rix_2018} are identified as candidate multiple star systems. This candidate sample is then cleaned of clusters, moving groups, and resolved triples by making cuts on the number of neighbors for each candidate pair. Finally, as described in ERH21, the chance alignment probability for each pair is estimated empirically using a mock catalog of chance alignments constructed from DR3 data. Only candidates with a $<10\%$ probability of being a chance alignment are kept in the catalog. The vast majority of pairs in the catalog have chance alignment probabilities much lower than $10\%$: the median and 95th percentile chance alignment probabilities are about $0.01\%$ and $2\%$, respectively.

In ERH21, the authors selected candidate pairs as those with component parallaxes consistent within 3 (or 6) sigma for angular separation (i.e.\ $\theta$) greater than (less than) 4 arcseconds as candidate binaries. They adopted a different threshold at $\theta < 4$ arcseconds because of the lower chance alignment rates and significantly underestimated parallax uncertainties at close angular separations. We loosen this criterion to 4 (or 8) sigma because we anticipate the possibility that the parallax uncertainties are more strongly underestimated for astrometric binaries in comparison to single stars. A less stringent cut also allows us to build a more extensive catalog, derive more accurate fits to the distributions of uncertainty-normalized parallax difference in each magnitude bin, and better characterize the tails of these distributions (see Section \ref{sec:calibration} for details).

 We focus our subsequent analysis on the wide pairs in which one component has an orbital or acceleration solution published in DR3. In these systems, a tertiary companion orbits an unresolved binary in a wide orbit. Some of these unresolved binaries have full orbital solutions from \textit{Gaia} DR3, while others only have acceleration solutions. Either of these solutions represents an improvement in fitting the motion of the source's photocenter on the sky when compared to the 5-parameter single star solution \citep{halbwachs_2023}. By selecting for primary sources with orbital or acceleration solutions, we curate a sample of triples that can be used to validate the uncertainties for these astrometric binaries, or for astrophysical studies of hierarchical multiplicity.
 
In the rest of this paper and in our published catalog, ``primary'' refers to Star 1, labeled as such because it has an astrometric binary solution; the distant tertiary is labeled as the ``secondary'' or Star 2. As we show below, the astrometric binary is usually, but not always, also the brighter of the two sources. Our final sample is publicly available, and includes 7,820 systems hosting unresolved binaries with astrometric solutions and 6,998 systems hosting unresolved binaries with acceleration solutions. There are 27 systems that fall into both categories. We describe the catalog in more detail in Appendix~\ref{sec:catalog}.

\subsection{Basic properties of the sample}

\begin{figure*}
    \centering
    \includegraphics[width=0.7\textwidth]{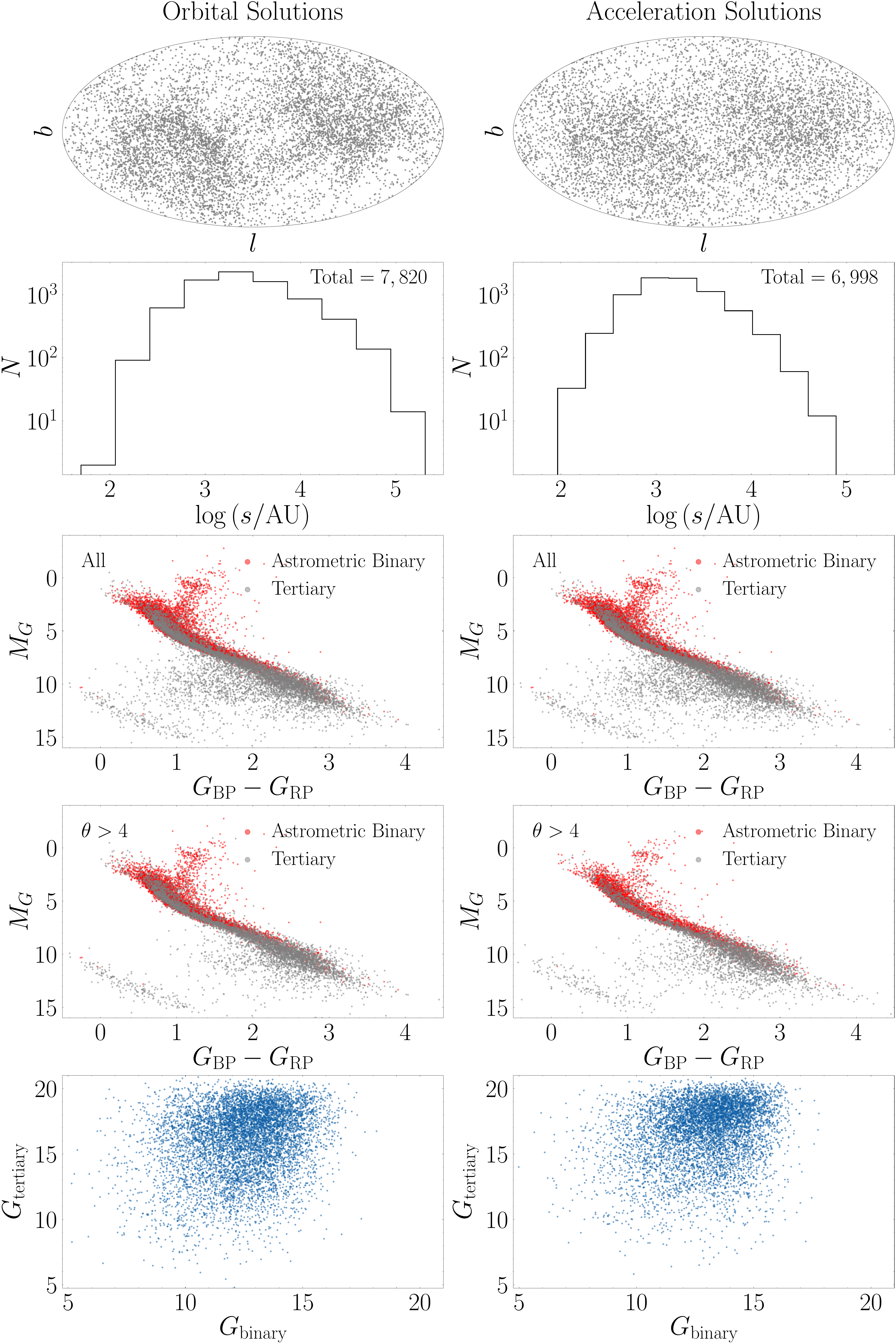}
    \caption{Basic properties of the hierarchical triple samples. The left and right columns display the properties of systems hosting astrometric binaries with orbital and acceleration solutions, respectively. Top row: Sky distribution in Galactic coordinates. Imprints of the \textit{Gaia} scanning law and the Galactic disk are visible in both columns. Second row: Distributions of projected physical separations of between the inner binary and resolved tertiary. The distributions fall off at small $s$ because close outer tertiaries are unresolved, and at large $s$ because wide bound companions are rare and difficult to distinguish from chance alignments. Third row: Color-magnitude diagrams. Unresolved astrometric binaries are plotted in red, and tertiaries are plotted in gray. While both the inner binaries and outer tertiaries primarily fall along the main sequence, some of the inner binaries fall on the red giant branch, and some of the tertiaries are white dwarfs. Fourth row: Same as the third row, except we only show systems with angular separations greater than 4 arcseconds. Most of the the sources falling between the main sequence and white dwarf track --- which are primarily sources with inaccurate colors due to blending --- are removed by this cut. Fifth row: Apparent $G$-band magnitudes of astrometric binaries and tertiaries. In both columns, the majority of the astrometric binaries fall between $G = 8$ and $G = 16$, reflecting the fact that most astrometric binary solutions published in \textit{Gaia} DR3 are for bright sources. The tertiaries are fainter on average, though we focus on the bright tertiaries in our subsequent analysis.}
    \label{fig:summary_plot}
\end{figure*}

We visualize the basic properties of the sample in Figure~\ref{fig:summary_plot}. The left and right columns correspond to hierarchical triples hosting unresolved binaries with orbital solutions and acceleration solutions, respectively. In the top row, we plot the sky distributions of the astrometric binaries in Galactic coordinates. In both cases, we observe structure arising due to the \textit{Gaia} DR3 scanning law, and detect hints of the Galactic plane. Imprints of the \textit{Gaia} scanning law are more evident in the sky distribution corresponding to the orbital solutions than in the sky distribution corresponding to the acceleration solutions. In the second row, we show the distributions of projected physical separations between the inner binary and outer tertiary. These distributions peak at $s \sim 1000$ AU, and span the entire range of projected physical separations studied in ERH21. At small separations ($s \lesssim 100$ AU), outer tertiaries are unresolved. At large separations ($s \gtrsim 10000$ AU), bound companions are rare and more difficult to distinguish from chance alignments. In the third row, we show \textit{Gaia} color-magnitude diagrams of both components. The unresolved astrometric binaries are plotted in red, while the tertiaries are plotted in gray. We find that both the inner binaries and outer tertiaries primarily fall along the main sequence, with a portion of the inner binaries falling on the red giant branch, and a small subset of the tertiaries falling on the white dwarf cooling track. We restrict the triples plotted on the color-magnitude diagrams to those with angular separations greater than 4 arcseconds in the fourth row. This cut removes most tertiary stars that fall below the main sequence due to inaccurate colors from blending \citep[e.g.][]{rybizki_2022}. Finally, in the fifth row, we visualize the distributions of apparent $G$-band magnitudes of astrometric binaries and tertiaries. The majority of the astrometric binaries fall between $G = 8$ and $G = 16$. These binaries are on average a few magnitudes brighter than their tertiary counterparts. This is because faint (i.e.\ $G \gg 14$) sources generally have astrometric errors that are too large to receive an astrometric binary solution in \textit{Gaia} DR3.

\begin{figure*}
    \centering
    \includegraphics[width=\textwidth]{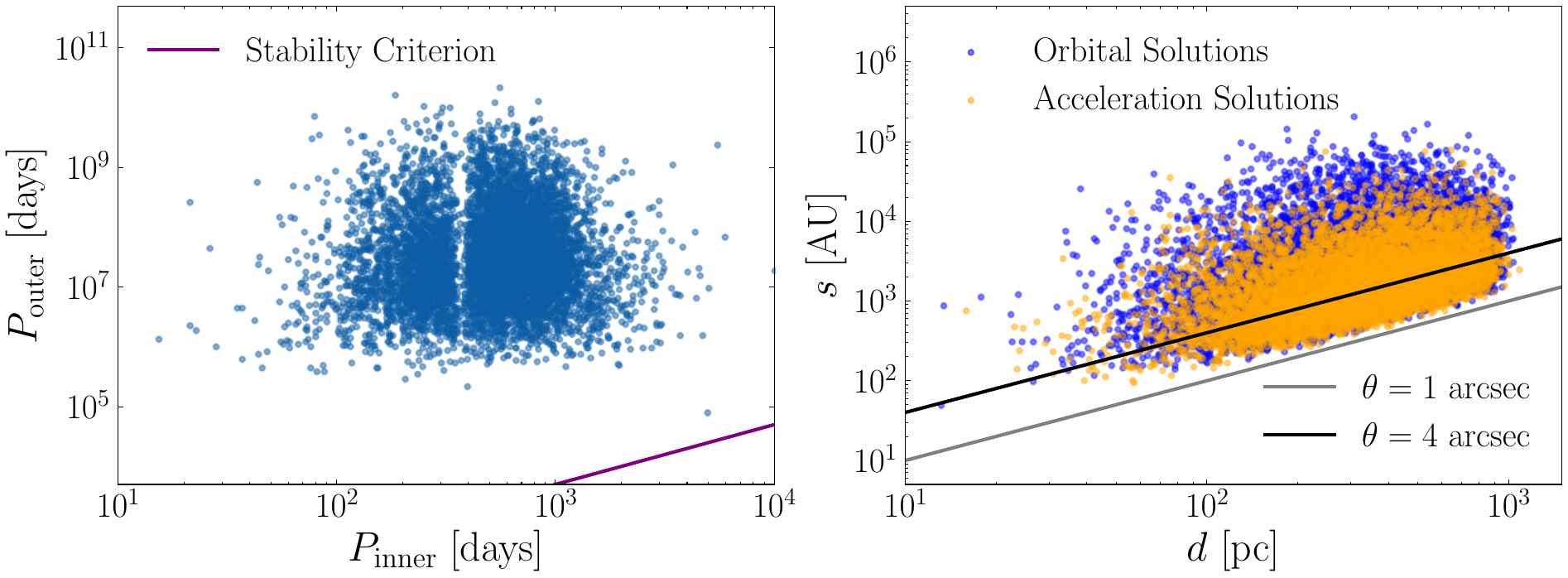}
    \caption{Left: Inner and outer orbital periods for systems in which the inner binary has an astrometric orbital solution, assuming a total mass of $2.5\,M_{\odot}$. A gap exists at $P_{\text{inner}} = 1$ year due to degeneracy with the parallax measurement. Besides $P_{\text{outer}}$ being $5$--$6$ orders of magnitude longer than $P_{\text{inner}}$ on average, no other correlation is observed. All systems lie well within the regime of stability ($P_{\text{outer}} > 5 P_{\text{inner}}$). Right: Projected physical separation as a function of distance for astrometric binaries with orbital (blue) and acceleration (orange) solutions. Most sources lie between $100$--$1000$ pc away and have projected physical separations between $100$--$10^5$ au. All systems have angular separation $> 1$ arcsec. Our analysis is focused on the systems with angular separation $> 4$ arcsec.}
    \label{fig:kepler_fig}
\end{figure*}

Assuming a total mass of $2.5\,M_{\odot}$, we use Kepler's third law to plot the orbital periods of the outer tertiaries against the orbital periods of the inner binaries for the systems with full orbital solutions in the left panel of Figure~\ref{fig:kepler_fig}. A gap in the distribution exists at an inner binary orbital period of $1$ year due to the degeneracy with the periodicity of the parallactic motion of the sources. On average, the orbital periods of the tertiary stars are $5$--$6$ orders of magnitude longer than the orbital periods of the inner binaries. Otherwise, no correlation is observed. The systems all lie within the regime of dynamical stability defined by the empirical relation $P_{\text{outer}} > 5 P_{\text{inner}}$ \citep[e.g.][]{tokovinin_2004}. We plot the angular separations of the triples as a function of distance in the right panel of Figure~\ref{fig:kepler_fig}, with the sources hosting astrometric binaries with orbital and acceleration solutions overlaid in blue and orange, respectively. The majority of sources lie between $100$ pc and $1$ kpc from the Sun and have projected physical separations between $100$--$10^5$ au. All systems have angular separation $> 1$ arcsec, with closer pairs missing from the sample due to the tertiaries degrading the quality of the astrometric binary solutions. We demarcate the region corresponding to angular separation $> 4$ arcsec with a solid black line. We only consider pairs wider than this limit in our subsequent analysis.

\section{Calibration of parallax uncertainties for unresolved astrometric binaries in \textit{Gaia} DR3} \label{sec:calibration}

Because the unresolved binary and the tertiary companion should have nearly the same distance, our catalog provides a straightforward method of validating the parallax uncertainties for astrometric binaries in \textit{Gaia} DR3. To this end, we calculate the uncertainty-normalized parallax difference between the unresolved binary and the tertiary companion, given by:

\begin{equation}
    \frac{\Delta \varpi}{\sigma_{\Delta \varpi}} = \frac{\varpi_{\text{binary}} - \varpi_{\text{tertiary}}}{\sqrt{\sigma_{\varpi_{\text{binary}}}^2 + \sigma_{\varpi_{\text{tertiary}}}^2}}\,.
\end{equation}

Here, $\Delta \varpi$ represents the parallax difference between the unresolved binary and the tertiary, while $\sigma_{\Delta \varpi}$ represents the uncertainty in that quantity. $\varpi_{\text{binary}}$ and $\varpi_{\text{tertiary}}$ are the measured parallaxes of the unresolved binary and tertiary, and $\sigma_{\varpi_{\text{binary}}}$ and $\sigma_{\varpi_{\text{tertiary}}}$ are the corresponding uncertainties.\footnote{Both here and in all other places in the text, $\varpi_{\text{tertiary}}$ represents the parallax uncertainty of the tertiary after it has been inflated according to the fitting function $f(G)$ given in Equation 16 of ERH21.}

We now introduce the parallax underestimate factor $f$, generally defined such that: 

\begin{equation}
\label{eq:f}
    \sigma_{\varpi,\,{\text{true}}} = f \times \sigma_{\varpi,\,{\text{reported}}}\,,
\end{equation}

where $\sigma_{\varpi,\,{\text{true}}}$ is the true parallax uncertainty and $\sigma_{\varpi,\,{\text{reported}}}$ is the parallax uncertainty published in \textit{Gaia} DR3.

If the parallax uncertainties for the unresolved binaries are accurate, $\Delta \varpi / \sigma_{\Delta \varpi}$ should be distributed as a Gaussian with $\sigma = 1$. On the other hand, if those reported parallax uncertainties are underestimated, then we will observe a wider distribution instead. In addition, if there is an offset in the parallax zeropoint between the 5-parameter and astrometric binary
solutions, the distribution might not be centered on zero.

In our subsequent analysis, we only examine pairs with $\theta > 4$ arcsec. This is because ERH21 showed that sources with neighbors within 4 arcseconds often have more strongly underestimated parallax errors than isolated sources. Here, we are interested in estimating how much parallax errors are underestimated for normal astrometric binaries, most of which do not have a bright companion within 4 arcseconds. 

In this and all subsequent calculations, we also exclude hierarchical triples that are wide and nearby enough that their physical size might measurably contribute to the parallax difference. This condition corresponds to $\Delta \varpi_{\text{true}} / \sigma_{\Delta \varpi} > 0.05$, where (following ERH21) we define the approximate true parallax difference $\Delta{\varpi}_{\text{true}}$ as:

\begin{equation}
    \Delta \varpi_{\text{true}} = \frac{1}{206265} \text{ mas} \times \left(\frac{\theta}{\text{arcsec}}\right)\left(\frac{\varpi}{\text{mas}}\right)\,.
\end{equation}

Lastly, our catalog contains 73 pairs in which both components have an astrometric binary solution. We remove these systems, which are likely quadruple star systems, from the following analysis, but retain them in the published catalog for exploration in future work.

\subsection{Orbital solutions}
\label{sec:orbital_sols}

\begin{figure*}
    \centering
    \includegraphics[width=\textwidth]{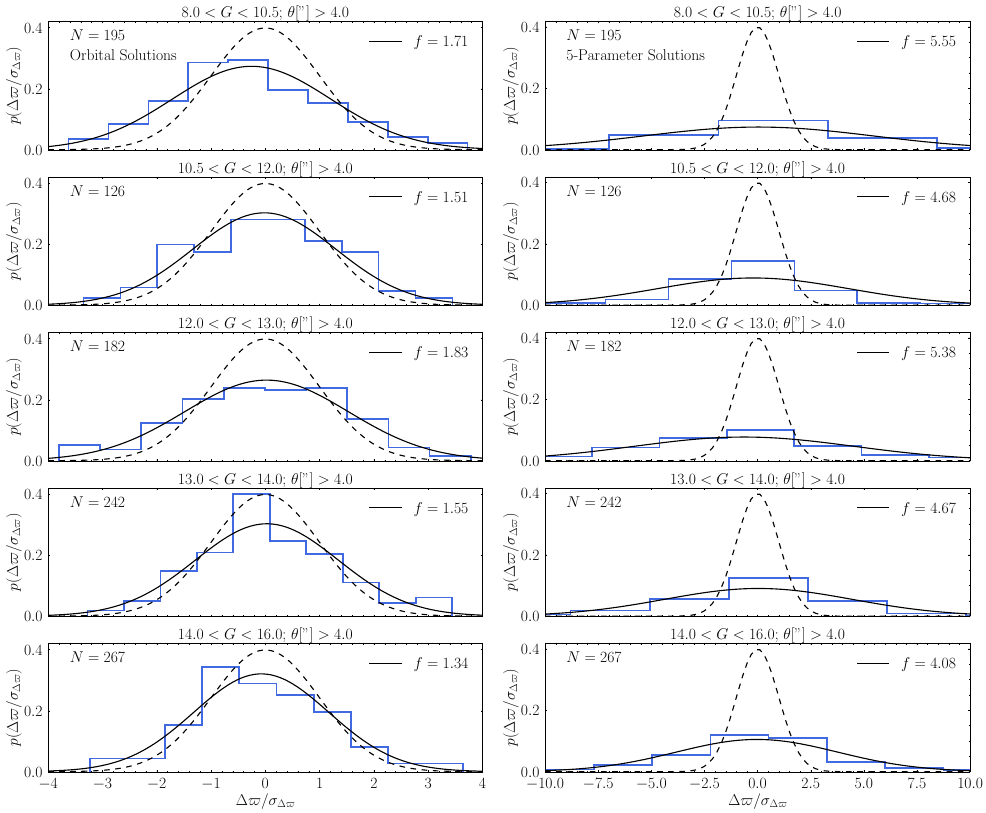}
    \caption{Left: Distributions of uncertainty-normalized parallax difference between the unresolved binary and the tertiary companion. Each panel shows a different bin of primary apparent magnitude. Because the unresolved binary and tertiary have essentially the same distance, this quantity should follow a Gaussian distribution with $\sigma$ = 1 (dotted line) if the \textit{Gaia} DR3 parallax uncertainties are accurate. Solid black lines show the best-fit truncated Gaussian distributions. We find that $f > 1$ in all cases, implying underestimated uncertainties. The uncertainties are underestimated more strongly for bright sources than for sources in the faintest magnitude bin. Right: Same as left column, but using the parallaxes from the single-star solutions rather than the binary solutions. The much broader distributions found in this case demonstrate that the binary solutions yield more accurate parallaxes than the single-star solutions for the same sources.}
    \label{fig:unpd}
\end{figure*}

We begin by considering hierarchical triples hosting unresolved astrometric binaries with orbital solutions. In Figure~\ref{fig:unpd}, we show distributions of $\Delta \varpi / \sigma_{\Delta \varpi}$ for these hierarchical triples at a range of primary apparent $G$-band magnitudes. In order for the distribution of $\Delta \varpi / \sigma_{\Delta \varpi}$ to be sensitive to the parallax errors of the astrometric binary solutions, these need to be comparable to or larger than the parallax errors of the single-star solutions. If the primary has magnitude $G < 10.5$, we allow the secondary to be as faint as $G = 13$; otherwise, we require that the secondary is not fainter than the primary by more than 0.5 magnitudes. This is reasonable because $\sigma_\varpi(G)$ is fairly flat at $G < 13$ for $5$-parameter solutions \citep{lindegren_2021}. 

Blue histograms in Figure~\ref{fig:unpd} show the observed distributions. We show five bins of apparent $G$-band magnitude of varying width, chosen so that a comparable number of systems land in each bin. Also shown in Figure~\ref{fig:unpd} are truncated Gaussian fits to the data. Because pairs with $| \Delta \varpi | / \sigma_{\Delta \varpi} < 4$ do not enter the sample, it is necessary to account for this truncation of the distribution. We assume the observed values of  $\Delta \varpi/ \sigma_{\Delta \varpi}$ are drawn from a distribution defined as:

\begin{equation}
\label{eq:truncated_normal}
 p\left(\frac{\Delta \varpi}{\sigma_{\Delta \varpi}}\right) = \frac{{\displaystyle\frac{1}{\sigma}}\, \varphi\left({\displaystyle\frac{\Delta \varpi / \sigma_{\Delta \varpi} - \mu / \sigma_{\Delta \varpi}}{\sigma}}\right)}{\Phi\left({\displaystyle\frac{b - \mu / \sigma_{\Delta \varpi}}{\sigma}}\right) - \Phi\left({\displaystyle\frac{a - \mu / \sigma_{\Delta \varpi}}{\sigma}}\right)}
\end{equation}

for $a \leq x \leq b$, and $0$ otherwise. Here, $\sigma^2$ is the variance of the truncated normal distribution, $\mu$ represents the parallax zeropoint offset between the astrometric binary and 5-parameter single star solutions, $\varphi$ is the probability density function of a unit-variance normal distribution,

\begin{equation}
    \varphi(z) = \frac{1}{\sqrt{2\pi}} \exp\left(-\frac{z^2}{2}\right)\,,
\end{equation}

$\Phi$ is its cumulative distribution function,

\begin{equation}
    \Phi(x) = \frac{1}{2} \left[1 + {\rm erf} \left(\frac{x}{\sqrt{2}}\right)\right]\,,
\end{equation}

and our bounds are $(a, b) = (-4, 4)$. We express $\sigma$ as a function of the parallax uncertainty underestimate factor $f$:

\begin{equation}
    \sigma = \sqrt{\frac{(f \times \sigma_{\varpi_{\text{binary}}})^2 + \sigma_{\varpi_{\text{tertiary}}}^2}{\sigma_{\varpi_{\text{binary}}}^2 + \sigma_{\varpi_{\text{tertiary}}}^2}}\,.
\end{equation}

This accounts for the fact that only one component is expected to have underestimated parallax uncertainties. The log-likelihood for a set of uncertainty-normalized parallax differences is then given by

\begin{equation}
\label{eq:likelihood}
    \ln L = \sum_i \ln p\left((\Delta \varpi / \sigma_{\Delta \varpi})_i\right)\,,
\end{equation}

where the sum is calculated over all systems in the set. In summary, the free parameters are $f$ and $\mu$. The binning of the data is only used for visualization purposes, and is not used in the fitting of the continuous truncated Gaussian distribution. We verified using mock data that this approach yields accurate and unbiased constraints on $f$ and $\mu$.

We assume broad, flat priors on the model parameters. Then, for each panel of Figure~\ref{fig:unpd}, we use ensemble MCMC sampling \citep[\texttt{emcee};][]{emcee_2013} with $64$ walkers and $500$ steps per walker to sample from Equation~\ref{eq:likelihood} and derive the values of $f$ and $\mu$ that best describe the truncated Gaussian distribution. The best-fit truncated Gaussian (using the median values of the mean and variance in each magnitude bin) is plotted with a solid line, and the corresponding value of $f$ is shown in the legend. For comparison, we also plot a Gaussian with $\sigma = 1$ and $\mu = 0$. From Figure~\ref{fig:unpd}, it is clear that the parallax uncertainties are underestimated (i.e.\ $f > 1$) in all magnitude bins. We also find that the parallaxes are underestimated more strongly for the bright sources than for the sources in the faintest bin, with the value of $f$ rising from $\approx 1.3$ at $G \gtrsim 14$ to $\approx 1.7$ at $G \lesssim 14$. 

\subsubsection{Cuts based on the \texttt{goodness\_of\_fit} metric}

Next, to check the effect of the \texttt{goodness\_of\_fit} \citep[$F_2$;][]{Wilson_Hilferty_1931} metric of the orbital solutions on our results, we restrict the dataset to unresolved astrometric binaries with ``good'' orbital solutions, with the cut based on the apparent magnitude of the primary:

\begin{equation}
\label{eq:good}
    \texttt{goodness\_of\_fit} <
    \cases{
    5 & for $G \geq 13$ \cr
    10 & for $G < 13\,.$ \cr
    }
\end{equation}

\begin{figure*}
    \centering
    \includegraphics[width=\textwidth]{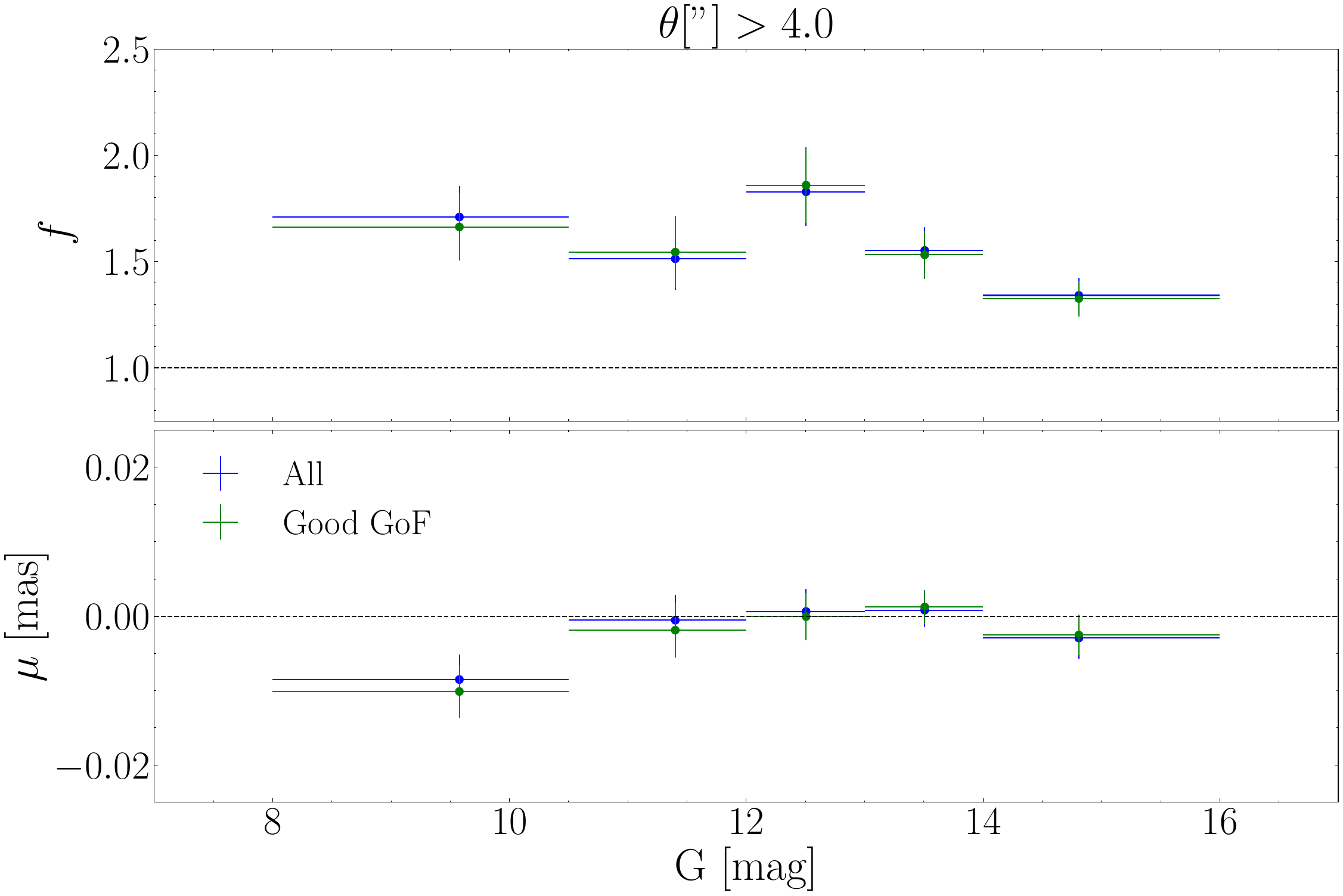}
    \caption{Parallax uncertainty underestimate factor (i.e.\ $f$, see Equation~\ref{eq:f}) and parallax zeropoint offset between single-star and astrometric binary solutions as a function of primary $G$-band apparent magnitude. We plot the best-fit parameters of the truncated Gaussians for both the full analysis from Figure~\ref{fig:unpd} and the analysis restricted to ``good'' astrometric binary solutions (see Equation~\ref{eq:good}). The horizontal error bars represent the widths of each bin, while the vertical error bars represent the standard deviations of the posterior samples in each bin. The parallax uncertainties are more strongly underestimated for brighter sources than fainter sources, with the largest underestimate factor of $f \approx 2$ occurring in the $12 < G < 13$ bin. We find that any offset in parallax zeropoint between the single-star and astrometric binary solutions is small ($\sim 0.01$ mas).}
    \label{fig:good_trend_fig}
\end{figure*}

We make this choice due to a discontinuity in the \texttt{goodness\_of\_fit} distribution at $G = 13$ for solutions in DR3; the point of the cut is to select only astrometric solutions with relatively good fits \citep[see e.g.][their Figure 10]{gaia_ns_population}. The best-fit values of $f$ and $\mu$ in each magnitude bin are compared to the values derived from the full analysis in Figure~\ref{fig:good_trend_fig}. Even after the cut on \texttt{goodness\_of\_fit}, we see similar trends in $f$ as in Figure~\ref{fig:unpd}, and the parallax uncertainties are still underestimated in each magnitude bin, indicating that our inferred values of $f$ are not driven by sources with poor \texttt{goodness\_of\_fit}. We find that, at separations of $\theta > 4$ arcseconds, the parallax uncertainties are most strongly underestimated in the $12 < G < 13$ bin, by almost a factor of two. ERH21 found a similar result for the 5-parameter solutions; the increase in the value of $f$ in this bin may be explained by various changes in the \textit{Gaia} data processing at $G \approx 13$ (see discussion in ERH21). We also determine that any offset in parallax zeropoint between the single-star and binary solutions is only on the order of $\sim 0.01$ mas.

We also perform our analysis on unresolved astrometric binaries with ``bad'' orbital solutions (i.e., those not passing the cut defined in Equation~\ref{eq:good}). There are 93 sources with apparent $G$-band magnitudes between $G = 9$ and $G = 14$ that have \texttt{goodness\_of\_fit} values that do not pass our cut. We find that the distribution of uncertainty-normalized parallax difference for these sources is best described by a truncated Gaussian with $f = 1.69 \pm 0.20$ and $\mu = 0.004 \pm 0.005$ mas. These values are consistent with our results from the analysis of the ``good'' sources to within $2\sigma$, reinforcing our conclusions that our inferred values of the parallax uncertainty inflation factor are not strongly dependent on \texttt{goodness\_of\_fit}.

\subsubsection{Comparison of orbital solutions with 5-parameter solutions}

\begin{figure*}
    \centering
    \includegraphics[width=\textwidth]{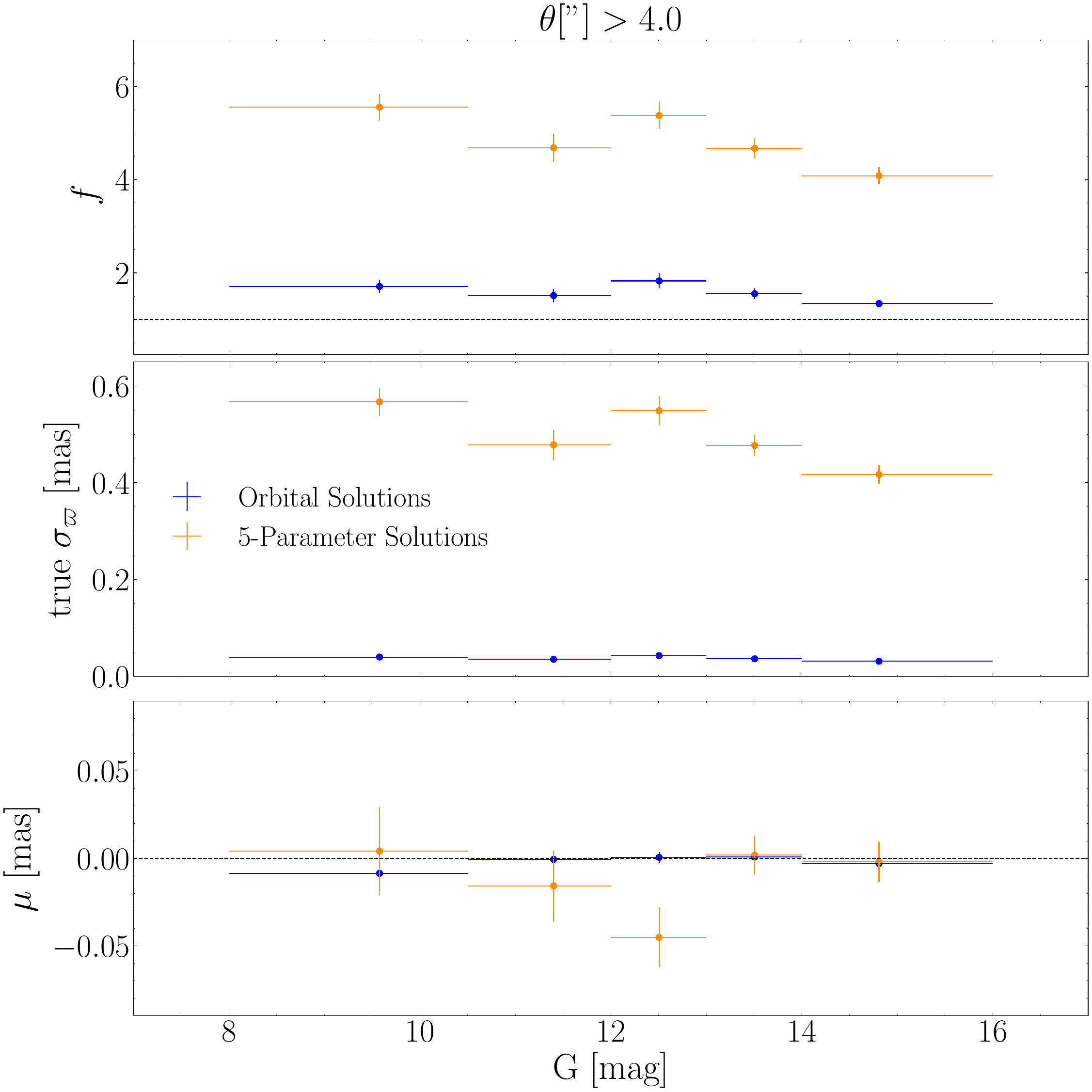}
    \caption{Same as Figure~\ref{fig:good_trend_fig}, except that we now compare the best-fit parameters from the analysis of the orbital solutions against the best-fit parameters from the analysis of the corresponding 5-parameter single-star solutions. In each magnitude bin, the value of $f$ is $2$--$3$ times larger when the 5-parameter solutions are used in place of the binary solutions, implying that the binary solutions are superior to the single star solutions. Since the reported parallax uncertainties are also significantly larger for these sources' 5-parameter solutions, the true parallax uncertainties are typically $\approx 10$ times larger for the single star solutions than for the orbital solutions in each magnitude bin. We find that the parallax zeropoint offset between the 5-parameter solutions of the astrometric binary and the tertiary is consistent with zero in all bins except the $12 < G < 13$ magnitude bin, where it is about $-0.05$ mas.}
    \label{fig:five_par_trend_fig}
\end{figure*}

We now repeat our analysis, except that for each unresolved binary, we replace the parallax from the astrometric binary solution with the parallax from the 5-parameter single star solution. In this case, we do not truncate the fitted Gaussian distribution, as no cuts on the consistency of the parallaxes from the 5-parameter solutions were used in constructing the catalog. We show the resulting distributions of uncertainty-normalized parallax difference between the unresolved binary and the tertiary companion in the right column of Figure~\ref{fig:unpd}. We find that this change results in the parallax uncertainties being severely underestimated in each magnitude bin, demonstrating that the astrometric orbital solutions provide much more accurate parallax constraints than the 5-parameter solutions for the same sources. The best-fit values of $f$ and $\mu$ in each magnitude bin are compared to the values derived from the analysis of the binary solutions in Figure~\ref{fig:five_par_trend_fig}. We also compare the median true parallax uncertainties (i.e.\ after inflation by $f$, see Equation~\ref{eq:f}) of the two kinds of solutions in each magnitude bin. At separations of $\theta > 4$ arcseconds, the parallax uncertainty underestimate factor is $2$--$3$ times larger when the 5-parameter solutions are used in place of the orbital solutions in each magnitude bin. Moreover, the reported parallax uncertainties are significantly larger for these sources' 5-parameter solutions, so the true parallax uncertainties are typically $\approx 10$ times larger for the 5-parameter solutions than for the orbital solutions in each magnitude bin. The parallax zeropoint offset between the 5-parameter solutions of the astrometric binary and the tertiary is consistent with zero in all bins except the $12 < G < 13$ magnitude bin, where it is about $-0.05$ mas.

\subsection{Acceleration solutions}
\label{sec:acc_sols}

\begin{figure*}
    \centering
    \includegraphics[width=\textwidth]{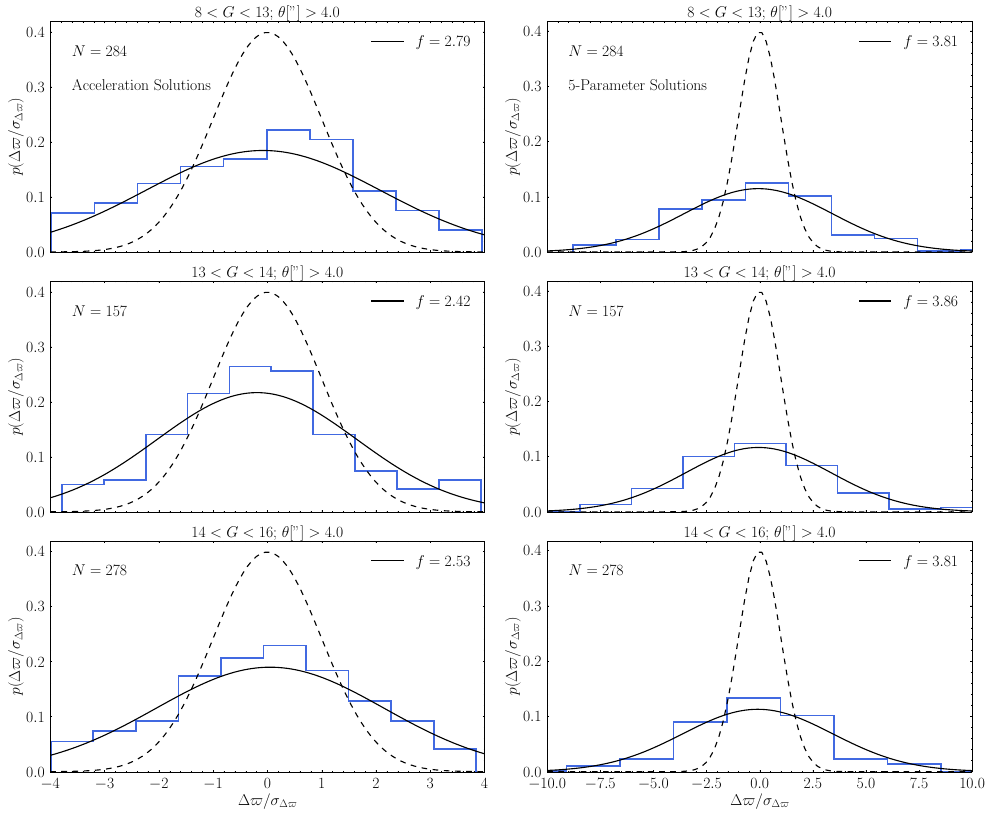}
    \caption{Same as Figure~\ref{fig:unpd}, but for unresolved binaries with acceleration solutions. As with the orbital solutions, we find that $f > 1$ in all cases, pointing towards underestimated parallax uncertainties. The value of the parallax uncertainty underestimate factor is similar in all magnitude bins. When 5-parameter solutions are used in place of acceleration solutions, the inferred values of $f$ are even larger, suggesting that the acceleration solutions still represent an improvement over single-star solutions.}
    \label{fig:acc_unpd}
\end{figure*}

We repeat the analysis in Section \ref{sec:orbital_sols} for systems hosting unresolved astrometric binaries with acceleration solutions. Once again, we only consider tertiary separations $\theta > 4$ arcseconds. We also reduce the number of $G$-band magnitude bins and expand the size of each magnitude bin so that enough triples exist in each bin to carry out a robust statistical analysis. We do not make any cuts on \texttt{goodness\_of\_fit}, as this metric is not very useful for acceleration solutions, which are rarely an accurate description of the sources' astrometric motions and thus generally have large \texttt{goodness\_of\_fit} values.

From Figure~\ref{fig:acc_unpd}, it is clear that the parallax uncertainties for the acceleration solutions are underestimated (i.e.\ $f > 1$) in all magnitude bins. Moreover, the parallax uncertainty underestimate factor is larger at all magnitudes than for the orbital solutions. We find that the value of the parallax uncertainty underestimate factor is $\approx 2.5$ in all magnitude bins. We also find that the value of $f$ is larger for sources with $9$-parameter acceleration solutions than for sources with $7$-parameter acceleration solutions (see Appendix \ref{sec:extra}).

\subsubsection{Comparison of acceleration solutions with 5-parameter solutions}

\begin{figure*}
    \centering
    \includegraphics[width=\textwidth]{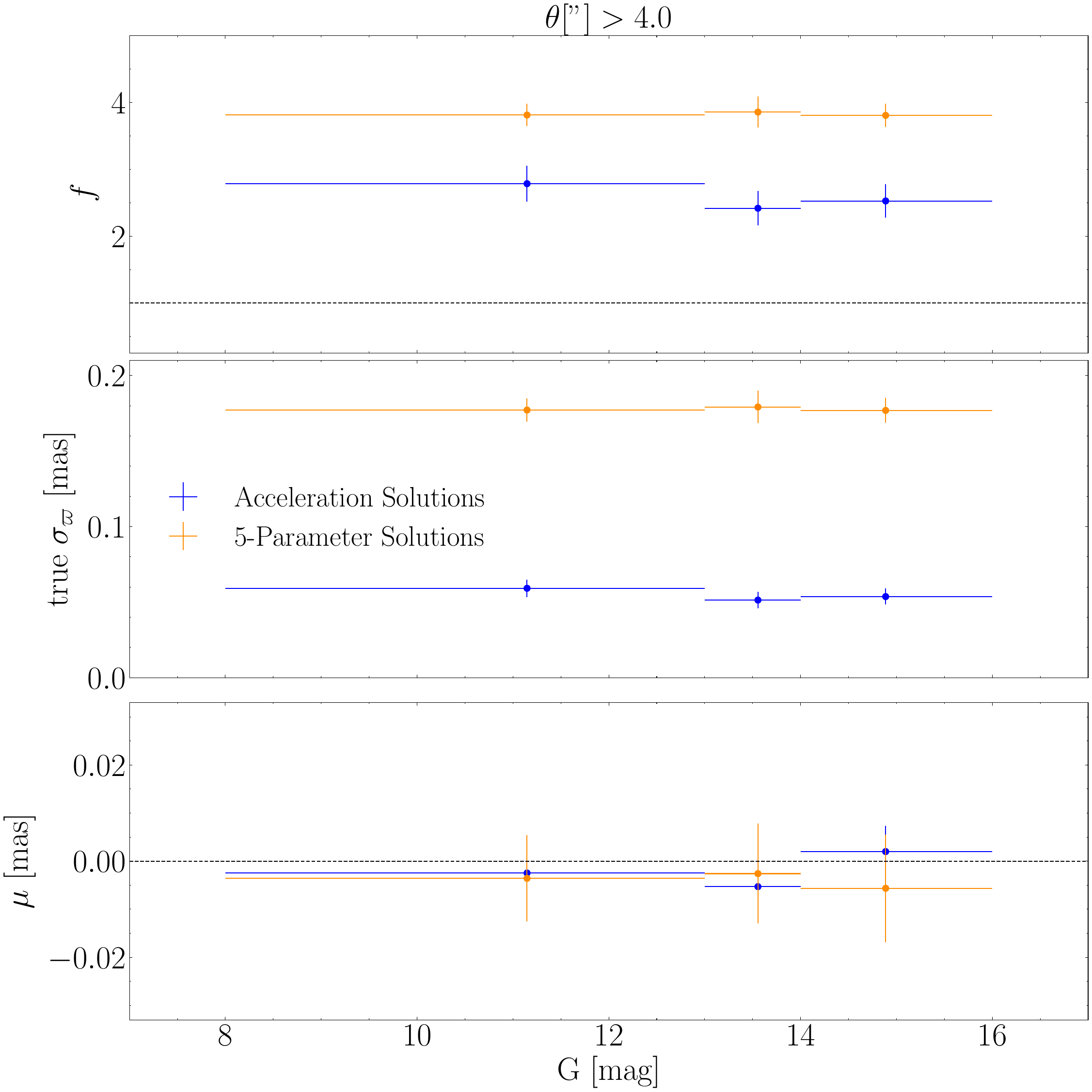}
    \caption{Same as Figure~\ref{fig:five_par_trend_fig}, except that we now compare the best-fit parameters from the analysis of the acceleration solutions against the best-fit parameters from the analysis of the corresponding 5-parameter single-star solutions. In each magnitude bin, the value of $f$ is $\approx 1.5$ times larger when the 5-parameter solutions are used in place of the acceleration solutions, implying that the acceleration solutions are superior to the single star solutions. Since the reported parallax uncertainties are also somewhat larger for these sources' 5-parameter solutions, the true parallax uncertainties are typically $\approx 3$ times larger for the single star solutions than for the acceleration solutions in each magnitude bin. We find that the parallax zeropoint offset between the astrometric binary and single star solutions is $\lesssim 0.01$ mas in all bins. The parallax zeropoint offset between the 5-parameter solutions of the astrometric binary and the tertiary is also $\lesssim 0.01$ mas in all bins. }
    \label{fig:acc_trend_fig}
\end{figure*}

We repeat the analysis using 5-parameter solutions, as we did for the orbital solutions. We show the resulting distributions of uncertainty-normalized parallax difference in the right column of Figure~\ref{fig:acc_unpd}. As with the orbital solutions, we find that this change results in the parallax uncertainties being more severely underestimated in each magnitude bin, implying that the acceleration solutions are an improvement compared to the 5-parameter solutions, though still imperfect. The best-fit values of $f$ and $\mu$ in each magnitude bin are compared against the values derived from the analysis of the acceleration solutions in Figure~\ref{fig:acc_trend_fig}. We also compare the median true parallax uncertainties of the two kinds of solutions in each magnitude bin. We find that, at separations of $\theta > 4$ arcseconds, the parallax uncertainty underestimate factor is $\approx 1.5$ times worse when the astrometric solutions are replaced by the 5-parameter solutions in each magnitude bin. Furthermore, the reported parallax uncertainties are somewhat larger for these sources' 5-parameter solutions, so the true parallax uncertainties are typically $\approx 3$ times larger for the 5-parameter solutions than for the acceleration solutions in each magnitude bin. We find that the parallax zeropoint offset between the astrometric binary and single star solutions is $\lesssim 0.01$ mas in all bins, and that the parallax zeropoint offset between the 5-parameter solutions of the astrometric binary and the tertiary is $\lesssim 0.01$ mas in all bins as well. 

\subsection{Prescription to inflate $\sigma_{\varpi}$}

We list our inferred parallax uncertainty inflation factors for orbital and acceleration solutions as a function of apparent magnitude in Table \ref{tab:prescription}. We also provide the derived parallax zeropoint offsets between the astrometric binary and single star solutions in each magnitude bin. The data summarized here are also plotted in blue in Figures \ref{fig:good_trend_fig} and \ref{fig:acc_trend_fig}. The best-fit values and uncertainties reported for the $\sigma_{\varpi}$ underestimate factors are derived from the median and standard deviation of the posterior samples, respectively. We recommend inflating the parallax uncertainties reported in DR3 for orbital and acceleration solutions by these factors.

\begin{deluxetable*}{cccc}
\tablecaption{Parallax uncertainty inflation factors and parallax zeropoint offsets relative to single star solutions for astrometric binary solutions in \textit{Gaia} DR3. The zeropoint offsets should be subtracted from the astrometric binary solutions to bring them to the same scale as the 5-parameter solutions. \label{tab:prescription}}
\tablehead{\colhead{Solution Type} & \colhead{$G$-band Apparent Magnitude} & \colhead{$\sigma_{\varpi}$ Underestimate Factor} & \colhead{Zeropoint Offset (mas)} \\
\colhead{(1)} & \colhead{(2)} & \colhead{(3)} & \colhead{(4)}}
\startdata
Orbital Solutions ($\theta > 4''$) & $8.0$--$10.5$ & $1.71 \pm 0.15$ & $-0.009 \pm 0.003$ \\
& $10.5$--$12.0$ & $1.51 \pm 0.15$ & $-0.001 \pm 0.003$ \\
& $12.0$--$13.0$ & $1.83 \pm 0.16$ & $0.001 \pm 0.003$ \\
& $13.0$--$14.0$ & $1.55 \pm 0.11$ & $0.001 \pm 0.002$ \\
& $14.0$--$16.0$ & $1.34 \pm 0.08$ & $-0.003 \pm 0.003$ \\
\hline
Acceleration Solutions ($\theta > 4''$) & $8.0$--$13.0$ & $2.79 \pm 0.27$ & $-0.002 \pm 0.004$ \\
& $13.0$--$14.0$ & $2.42 \pm 0.26$ & $-0.005 \pm 0.004$ \\
& $14.0$--$16.0$ & $2.53 \pm 0.25$ & $0.002 \pm 0.005$ \\
\enddata
\end{deluxetable*}

\subsection{Uncertainties of other astrometric parameters}

Astrometric binary solutions published in DR3 have 7, 9, 12, or 15 parameters, depending on solution type \citep{halbwachs_2023, arenou_2023}. Our analysis has focused on parallax uncertainties, which are the most straightforward to validate using wide hierarchical triples hosting unresolved binaries. However, it is quite likely that other parameters of the astrometric solutions (positions, proper motions, Thiele-Innes elements, etc.) also have underestimated uncertainties. In the absence of constraints on uncertainties of specific parameters, it seems reasonable to assume that all of the parameters' uncertainties are underestimated by a similar factor. We thus advocate inflating the uncertainties of other parameters of astrometric binary solutions by the same factors reported in Table \ref{tab:prescription}, though we emphasize that uncertainties of other parameters have not yet been systematically validated.\footnote{The proper motion differences between wide pairs in our catalog are dominated by the real orbital motion of the sources rather than the uncertainties in the measurements of their proper motions. Hence, it is not possible to perform a similar analysis using proper motions in place of parallaxes.}

\subsection{Possible reasons for the underestimated uncertainties}
\label{sec:reasons}

In Appendix \ref{sec:extra}, we explore trends with orbital period and astrometric solution significance. We find no strong trends in $f$ with either of these quantities. We briefly consider potential explanations for the underestimated uncertainties.

For the acceleration solutions, it is not surprising that $f$ is always large, because constant or linearly varying acceleration is rarely a good model for the photocenter's true motion. For binaries with orbital periods shorter than the observing baseline, the orbital model should be accurate in most cases. However, uncertainties could be preferentially underestimated for binaries if the astrometric measurements do not exactly trace the photocenter \citep[e.g.][]{Lindegren_2022}, or if the linearized uncertainty estimates used in fitting are not appropriate. \citet{halbwachs_2023} do not specify how uncertainties were calculated for orbital solutions, but we note that calculating uncertainties for these solutions is nontrivial. This is because three parameters are fit via nonlinear optimization, while the remaining nine parameters are solved for by linear regression. Finally, we note that the epoch-level uncertainties used in calculating the binary solutions are different from those used in calculating 5-parameter solutions, with the former having been empirically inflated and filtered for outliers \citep[][their Figure 3]{holl_2023}. This can be expected to lead to different uncertainties on the astrometric parameters at fixed brightness.

\section{Summary and Discussion} \label{sec:summary}

We have constructed a publicly available catalog of hierarchical triples in \textit{Gaia} DR3 and have used it to empirically validate the reported parallax uncertainties for unresolved binaries with astrometric or acceleration solutions. Our main results are as follows:

\begin{itemize}
    \item The full catalog contains 7,820 unresolved binaries with orbital solutions and 6,998 unresolved binaries with acceleration solutions. We construct this catalog by applying the methods from ERH21 to orbital and acceleration solutions from \textit{Gaia} DR3. These systems lie within $1$ kpc from the Sun and are distributed all across the sky. They have tertiary separations ranging from a few 10s of au to 1 pc, spanning the entire range of projected physical separations probed in ERH21 (Figures~\ref{fig:summary_plot} and~\ref{fig:kepler_fig}). 
    \item We use this sample to calibrate the parallax uncertainties published by \textit{Gaia} DR3 for astrometric binaries. This analysis uses the fact that the unresolved binary and tertiary have essentially the same distance and thus should generally have reported parallaxes that are consistent within their uncertainties. We show that these uncertainties are underestimated everywhere, but most severely for bright stars ($G \lesssim 14$), which includes most published astrometric binary solutions in DR3 (Figure~\ref{fig:unpd}). This conclusion remains unchanged when only sources with good reported \texttt{goodness\_of\_fit} values are considered (Figure~\ref{fig:good_trend_fig}). The underestimate factor is worse for acceleration solutions compared to orbital solutions (Figure~\ref{fig:acc_unpd}), and much larger if $5$-parameter solutions are used in place of the binary solutions (Figures \ref{fig:five_par_trend_fig} and \ref{fig:acc_trend_fig}).
    \item We provide an empirical prescription to inflate published parallax uncertainties (Table \ref{tab:prescription}). The parallax uncertainty underestimate factor for orbital solutions is $\approx 1.3$ for faint sources ($G \gtrsim 14)$, but rises to $\approx 1.7$ for brighter sources ($G \lesssim 14$). The parallax uncertainty underestimate factor for acceleration solutions is $\approx 2.5$ in all magnitude bins.
    \item \textit{Gaia} astrometric solutions hold promise for the discovery of dormant stellar-mass black holes, exoplanets, unique binaries, and other rare systems. We expect many more of these exciting objects to be discovered with the epoch astrometry and longer time baseline that will become available in \textit{Gaia} DR4. Our public catalog of hierarchical triples has straightforward applications in the validation of these scientific discoveries, since it is important to account for underestimated uncertainties on the parameters of these systems when analyzing their orbits.
\end{itemize}

%% Also note that the akcnowlodgment environment does not support long amounts of text. If you have a lot of people and institutions to acknowledge, do not use this command. Instead, create a new \section{Acknowledgments}.
\begin{acknowledgments}

We thank the referee for constructive feedback. We thank members of the Caltech/Carnegie dormant black hole task force for useful discussions. This research was supported by NSF grant AST-2307232. This work has made use of data from the European Space Agency (ESA) mission
{\it Gaia} (\url{https://www.cosmos.esa.int/gaia}), processed by the {\it Gaia}
Data Processing and Analysis Consortium (DPAC,
\url{https://www.cosmos.esa.int/web/gaia/dpac/consortium}). Funding for the DPAC
has been provided by national institutions, in particular the institutions
participating in the {\it Gaia} Multilateral Agreement.

\end{acknowledgments}


% \vspace{5mm}
% \facilities{}


\software{astropy \citep{2013A&A...558A..33A,2018AJ....156..123A}, \texttt{emcee} \citep{emcee_2013}}



\appendix

\section{Catalog Description}
\label{sec:catalog}

\begin{deluxetable*}{cc}
\tablecaption{Categorization of the 14,791 unique gravitationally bound wide pairs in our catalog. In total, there are 7,820 systems hosting astrometric binaries with orbital solutions, and 6,998 systems hosting astrometric binaries with acceleration solutions, with 27 systems falling into both categories. These 27 systems are part of a wider set of 73 systems which are quadruple star candidates. \label{tab:classification}}
\tablehead{\colhead{Pair Classification} & \colhead{Number of Pairs} \\
\colhead{(1)} & \colhead{(2)}}
\startdata
Orbital + Single-Star & 7,756 \\
Acceleration + Single-Star & 6,962 \\
Orbital + Orbital & 37 \\
Orbital + Acceleration / Acceleration + Orbital & 27 \\
Acceleration + Acceleration & 9 \\
\hline 
Total & 14,791 \\
\enddata
\end{deluxetable*}

The full catalog of gravitationally bound wide pairs will be hosted at CDS. It can also be accessed on Zenodo at \dataset[doi:10.5281/zenodo.12791391]{https://doi.org/10.5281/zenodo.12791391}. All columns from the \texttt{gaiadr3.gaia\_source} catalog are copied over for both the unresolved astrometric binary and its tertiary companion. Columns ending in “1” and “2” refer to the astrometric binary and putative tertiary, respectively. We also include columns \texttt{pairdistance} (angular separation $\theta$, in degrees), \texttt{sep\_AU} (projected separation $s$, in au), and \texttt{R\_chance\_align} ($\mathcal{R}$, see ERH21). Parallaxes, positions, and proper motions for sources that have astrometric solutions in the \texttt{gaiadr3.nss\_two\_body\_orbit} or \texttt{gaiadr3.nss\_acceleration\_astro} catalogs are replaced with their values from the respective catalogs. We apply the zeropoint correction of \citet{Lindegren_2021_zp} to all parallaxes. The parallax uncertainties of tertiaries which have single star solutions are inflated according to the fitting function $f(G)$ given in Equation 16 of ERH21. 

The sample contains 14,791 unique pairs. We break this sample into two tables: one containing 7,820 systems hosting astrometric binaries with orbital solutions, and the other containing 6,998 systems hosting astrometric binaries with acceleration solutions. There are 27 pairs containing one component with an acceleration solution and one with an orbital solution that are included in both tables. We categorize each pair according to the types of the solutions (i.e.\ orbital, acceleration, or single-star) that are provided for its individual components in \textit{Gaia} DR3. We present a detailed breakdown of the number of sources belonging to each category in Table~\ref{tab:classification}. In total, our catalog features 73 pairs in which both components have an astrometric binary solution. We consider these systems to be quadruple star candidates, and leave exploration of them to future work.

\section{Trends with Other Parameters}
\label{sec:extra}

\begin{figure*}
    \centering
    \includegraphics[width=\textwidth]{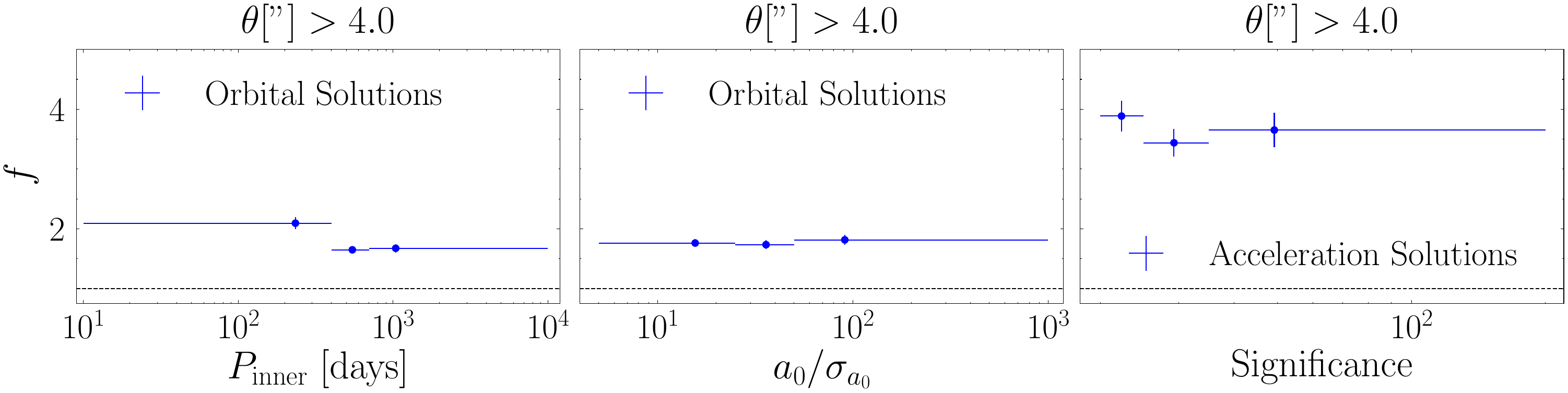}
    \caption{Left: Parallax uncertainty underestimate factor (i.e.\ $f$, see Equation~\ref{eq:f}) as a function of inner binary orbital period for systems hosting astrometric binaries with orbital solutions. The horizontal error bars represent the widths of each bin, while the vertical error bars represent the standard deviations of the posterior samples in each bin. Middle: Same as left panel, except as a function of signal-to-noise ratio of the astrometric orbit's semi-major axis. Right: Same as left panel, but for systems hosting inner binaries with acceleration solutions, binned by acceleration significance. There are no strong trends in $f$ with any of these parameters.}
    \label{fig:appendix_fig}
\end{figure*}

To investigate possible reasons for the underestimated parallax uncertainties found in the main text, we search for trends in the uncertainty underestimate factor $f$ with orbital period and with the significance of the astrometric solution. We first repeat our analysis for the systems hosting astrometric binaries with orbital solutions, binning by binary orbital period and $a_0 / \sigma_{a_0}$ (instead of $G$-band magnitude), respectively. Here, $a_0$ is the semi-major axis of the photocentric binary orbit and $\sigma_{a_0}$ is the error in that quantity; hence, $a_0 / \sigma_{a_0}$ represents the signal-to-noise ratio of the orbit. We again limit our analysis to pairs with separations greater than 4 arcsec, and exclude systems whose physical size could measurably contribute to the parallax difference. We also repeat our analysis for the systems hosting astrometric binaries with acceleration solutions, binning by the significance of the coefficients of the 7- and 9-parameter solutions (see \citet{halbwachs_2023} for details on how significance is defined). In each case, we divide the systems into three bins such that a roughly equal number of sources fall in each bin. We plot the parallax uncertainty underestimate factor $f$ as a function of each of these physical parameters in Figure~\ref{fig:appendix_fig}. We find no significant trend in $f$ with any of these parameters. 

Finally, we analyze the sources with 7-parameter and 9-parameter acceleration solutions separately. While we find that, on average, the value of $f$ is higher for the 9-parameter solutions ($f \approx 5$) than for the 7-parameter solutions ($f \approx 3$), our conclusion of a lack of trend in $f$ with significance holds for both types of solutions.

\newpage


\bibliography{bibliography}{}
\bibliographystyle{aasjournal}



\end{document}